\documentclass[sigconf,10pt]{acmart}

\renewcommand\footnotetextcopyrightpermission[1]{}

\AtBeginDocument{%
	\providecommand\BibTeX{{%
			\normalfont B\kern-0.5em{\scshape i\kern-0.25em b}\kern-0.8em\TeX}}}

\newcommand{\myName}{\textsf{Drone}}

\newcommand{\tabincell}[2]{\begin{tabular}{@{}#1@{}}#2\end{tabular}} 

\usepackage{algorithm}
\usepackage{algpseudocode}
\usepackage{graphicx}
\usepackage{amsmath,amsfonts,amsthm,bm}
\usepackage{diagbox}
\usepackage{subcaption}
\usepackage{hyperref,hyperxmp}
\usepackage{multirow}
\usepackage{bbding}
\usepackage{booktabs,siunitx}
\usepackage{tabularx}

\DeclareMathOperator*{\argmax}{arg\,max}
\theoremstyle{plain}

\theoremstyle{plain}

\theoremstyle{plain}

\theoremstyle{plain}

\theoremstyle{plain}

\usepackage{enumitem}
\setlist{nosep} 
\setlist{itemsep=1pt, topsep=3pt}

\usepackage{adjustbox} 
\begin{document}
\title[{\myName}: Dynamic Resource Orchestration for the Containerized Cloud]{Lifting the Fog of Uncertainties: Dynamic Resource Orchestration for the Containerized Cloud}
 \author{Yuqiu Zhang}
 \email{quincy.zhang@mail.utoronto.ca}
 \affiliation{%
   \institution{University of Toronto}
   \streetaddress{Department of Electrical and Computer Engineering}
   \country{Canada}
 }

 \author{Tongkun Zhang}
 \email{tongkun.zhang@mail.utoronto.ca}
 \affiliation{%
   \institution{University of Toronto}
   \streetaddress{Department of Electrical and Computer Engineering}
   \country{Canada}
 }

 \author{Gengrui Zhang}
 \email{gengrui.zhang@mail.utoronto.ca}
 \affiliation{%
   \institution{University of Toronto}
   \streetaddress{Department of Electrical and Computer Engineering}
   \country{Canada}
 }

 \author{Hans-Arno Jacobsen}
  \email{jacobsen@eecg.toronto.edu}
 \affiliation{%
   \institution{University of Toronto}
   \streetaddress{Department of Electrical and Computer Engineering}
   \country{Canada}
 }


\begin{abstract}
    
    The advances in virtualization technologies have sparked a growing transition from virtual machine (VM)-based to container-based infrastructure for cloud computing. From the resource orchestration perspective, containers' lightweight and highly configurable nature not only enables opportunities for more optimized strategies, but also poses greater challenges due to additional uncertainties and a larger configuration parameter search space. Towards this end, we propose {\myName}, a resource orchestration framework that adaptively configures resource parameters to improve application performance and reduce operational cost in the presence of cloud uncertainties. Built on Contextual Bandit techniques, {\myName} is able to achieve a balance between performance and resource cost on public clouds, and optimize performance on private clouds where a hard resource constraint is present. We show that our algorithms can achieve sub-linear growth in \textit{cumulative regret}, a theoretically sound convergence guarantee, and our extensive experiments show that {\myName} achieves an up to 45\% performance improvement and a 20\% resource footprint reduction across batch processing jobs and microservice workloads. 
  
\end{abstract}

\maketitle

\section{Introduction}\label{intro}

The last decade has witnessed the rise of cloud computing, a paradigm which revolutionized the entire IT industry~\cite{fox2009above,jonas2019cloud}. More recently, the rise of containerization technologies~\cite{solteszContainerbasedOperatingSystem2007} has further necessitated the prevalence of using the cloud to deploy large-scale analytical computing jobs or user-facing web applications thanks to containers' lightweight nature compared to their predecessor virtual machines (VMs). In terms of resource orchestration, containers also showcase much finer-grained controllability. For example, the smallest unit of resource configuration of a single container is in the magnitude of bytes for memory and millicores for CPU~\cite{docker_resource_constraints,k8s_resource}, while for VMs users are usually only able to select from a pre-configured family and instance type which differ from each other in the granularity of entire CPU cores and gigabytes for memory (e.g., <2vCPU, 4GiB Memory> for \texttt{m5.large} and <4vCPU, 16GiB Memory> for \texttt{m5.xlarge} instances on AWS EC2~\cite{aws_instance_types}). Furthermore, being smaller in size and having lower overhead to start, stop or schedule, containers are particularly suitable for deploying microservices, a new software architecture that breaks a monolithic application down to a set of loosely coupled components for independent and faster deployment and iteration~\cite{luo2021characterizing,eismann2020microservices}. 

Despite the benefits, due to additional fluidity they bring to the cloud, the wide adoption of containers also introduces greater \emph{cloud uncertainties}, which has already been an active research area~\cite{kabirUncertaintyawareDecisionsCloud2022}. A major uncertainty that significantly influences resource orchestration decisions is \textit{workload uncertainty}~\cite{kabirUncertaintyawareDecisionsCloud2022,luo2022power,dean2013tail}. Should a resource orchestration strategy fail to adjust to workload traffic in time, resource \emph{under-} or \emph{over-provisioning} might be caused, which results in service level agreement (SLA) violations or higher operational cost. Both cases are cost-inefficient and are hence undesirable. Consequently, numerous studies~\cite{luo2022power,al2017accrs,ghobaei2018autonomic,zhang2019mark,qiu2020firm,rzadca2020autopilot,kannan2019grandslam,baarzi2021showar} have proposed more intelligent approaches to adapt resource orchestration policies to workload changes. However, other uncertainties, though sometimes minor, can also become non-negligible factors that affect application performance or operational cost. Examples include existing cloud resource usage, inter-container traffic patterns, and the underlying computing framework on which the application runs~\cite{shi2015clash}. The shared nature of the cloud along with volatile incentives promoted by public cloud vendors (e.g., Spot and Burstable Instances on AWS~\cite{aws_spot,aws_burstable}) further makes the resource-performance/cost relationship non-structural and time-variant. Failing to consider these uncertainties may also lead to inefficient resource allocation policies, as we show in Section~\ref{obs}.

To address resource orchestration challenges induced by cloud uncertainties, we design {\myName}, an online resource orchestration framework for the containerized cloud. The idea is to progressively optimize a containerized application’s resource configuration over its lifespan with minimum manual intervention and the often-costly explicit workload profiling phase. At its core, {\myName} is built upon recent advances in Gaussian process-based contextual bandits~\cite{krause2011contextual}. By encompassing time-variant cloud uncertainties as contextual parameters, {\myName} follows an iterative procedure to continuously refine resource configurations based on the previous context-action pairs and collected performance metrics. Assuming a minimal structural relationship between application performance and resource configurations, the power of such a non-parametric model makes {\myName} versatile across a diverse range of cloud environments and adaptable to various application types and workloads. Specifically, we examine two settings within a shared cloud infrastructure: a) \emph{public cloud}, where computational resources can be effectively considered unlimited and {\myName} demonstrates adeptness in striking an efficient balance between performance and cost, and b) \emph{private cloud}, where there exists a stringent cap on computational resources and {\myName} proves capable of optimizing application performance within these resource constraints. {\myName} is also theoretically sound in both settings since it achieves a sublinear growth of cumulative regret, meaning that the algorithm converges fast with respect to its running time. 

We evaluate {\myName} by deploying various applications on our cloud-hosted Kubernetes cluster using {\myName} as an integrable resource orchestrator. Our extensive experimental analysis, employing realistic workloads, demonstrates {\myName}'s superior performance compared to alternative solutions in several respects. First, for recurring analytical jobs for which bandit-based approaches have been shown to be efficient~\cite{alipourfard2017cherrypick,liu2019accordia}, {\myName} exhibits further improvement in performance by accounting for a broader spectrum of cloud uncertainties, coupled with its adherence to resource constraints in the private cloud environment. Second, for user-facing microservices where workload variability is more ad-hoc and no explicit profiling phase is available, {\myName} also achieves a 37\% improvement on P90 latency compared to state-of-the-art alternatives, a result further amplified by our bespoke enhancements over the standard bandit optimization procedure, including a sliding window-based data sampler, empirically optimized starting point selection and latency-aware scheduling mechanisms. To the best of our knowledge, {\myName} is the first work to harness the potential of resource allocation in a containerized cloud using bandit-based approaches. It showcases superior adaptability across diverse settings in comparison to the preceding VM-based efforts. To sum up, we make the following contributions in this paper:

\begin{enumerate}
    \item Through comprehensive experimental analysis, we validate the non-structural performance-resource relationship and the significant influence of uncontrollable time-variant environment variables (the cloud uncertainties) on application performance under multiple cloud scenarios.
    \item Leveraging recent advances in bandit algorithms, we design {\myName}, a general-purpose online resource orchestration framework for container-based cloud systems. {\myName} progressively optimizes the performance-cost tradeoff in public cloud environments, while maintaining strict adherence to resource constraints in resource-limited private clouds. In both cases, {\myName} theoretically exhibits a fast convergence rate, guaranteeing its performance.
    \item We implement {\myName} as a customized resource orchestrator on top of Kubernetes. Using realistic cloud workloads, we show through extensive experiments that {\myName} outperforms state-of-the-art alternatives in terms of application performance, cost efficiency and resource constraint compliance.
\end{enumerate}

\section{Background and Related Work}\label{relatedwork}

\subsection{Cloud Resource Orchestration\label{sec:rw_cloud}}

\begin{table*}[ht]
	\renewcommand{\arraystretch}{1.1}
	\centering
 \begin{adjustbox}{width=\linewidth}
	\begin{tabular}{c|c|c|c|c|c|c|c|c}
		\toprule
		\hline
		{\textbf{Framework}} & {\tabincell{c}{\textbf{Application}}} & {\tabincell{c}{\textbf{Computing} \\\textbf{Platform}}}  & {\tabincell{c}{\textbf{Optimization} \\\textbf{Objective}}} & {\tabincell{c}{\textbf{Acquisition} \\\textbf{Function}}} & {\tabincell{c}{\textbf{Cloud} \\\textbf{Uncertainties} \\\textbf{(contexts)}}} & {\tabincell{c}{\textbf{Resource} \\\textbf{Limits} \\\textbf{(safety)}}} & {\tabincell{c}{\textbf{Workload}}} & {\tabincell{c}{\textbf{Convergence} \\\textbf{Guarantee}}}\\
		\hline
		{Dremel~\cite{zhao2022dremel}} & {DB Tuning} & {-} & {DB IOPS} & {UCB} & {\XSolidBrush} & {-} & {DB queries} & {\XSolidBrush}\\
            \hline
		{CGPTuner~\cite{cereda2021cgptuner}} & {DB Tuning} & {-} & {\tabincell{c}{Performance \\Improvement}} & {GP-Hedge} & {\tabincell{c}{Workload \\ only}} & {-} & {\tabincell{c}{Recurring\\DB queries}} & {\XSolidBrush}\\
		\hline
		{Cherrypick~\cite{alipourfard2017cherrypick}} & {\tabincell{c}{VM config.\\selection}} & {VM} & {Customized Cost} & {EI} & {\XSolidBrush} & {\XSolidBrush}& {\tabincell{c}{Recurring \\analytical jobs}} & {\XSolidBrush}\\
            \hline
            {Accordia~\cite{liu2019accordia}} & {\tabincell{c}{VM config.\\selection}} & {VM} & {Customized Cost} & {GP-UCB} & {\XSolidBrush} & {\XSolidBrush}& {\tabincell{c}{Recurring \\analytical jobs}} & {\checkmark}\\
            \hline
            {RAMBO~\cite{li2021rambo}} & {\tabincell{c}{Resource \\ orchestration}} & {Container} & {Customized Cost} & {SMSego} & {\XSolidBrush} & {\XSolidBrush}& {Microservices} & {\XSolidBrush}\\
            \hline
            {\myName} & {\tabincell{c}{Resource\\ orchestration}} & {Container} & {\tabincell{c}{Performance-cost \\ tradeoff (public cloud)\\Performance \\opt. (private cloud)}} & {GP-UCB} & {\checkmark} & {\checkmark}& {General} & {\checkmark}\\
	\hline
 		\bottomrule
	\end{tabular}
 \end{adjustbox}
 	\caption{Computer systems studies using bandit algorithms.}
	\label{table:comparison}
\end{table*}

Intelligent resource orchestration on the cloud has long been an active research area which can be categorized as follows based on the underlying techniques adopted. 


\noindent \textbf{Heuristic-based Approaches.} A simple yet practically effective resource orchestration choice is based on heuristics. They are usually intuitive and easy to implement and hence are widely adopted in industrial solutions~\cite{K8sauto,awsauto,gceauto,msauto}. For example, the default container autoscalers in Kubernetes~\cite{K8sauto} include \emph{Horizontal Pod Autoscaler (HPA)} and \emph{Vertical Pod Autoscaler (VPA)}, both of which follow a rule-based scaling policy. Such policies enable cloud tenants to define thresholds for interested metrics according to which the system performs autoscaling. However, setting appropriate thresholds for such metrics is a non-trivial task. The optimal values are often application-specific and require expert knowledge from the developer or system administrator. Therefore, such heuristic approaches can hardly generalize across various cloud applications and often involve significant manual efforts.


\noindent \textbf{Model-based analytical approaches.} Another line of work establishes analytical models to encapsulate the relationship between performance objectives and resource orchestration decisions. 
The problem is thus often modelled as an optimization problem and certain assumptions are usually made on the problem structure (e.g., linearity and convexity) so that theoretical properties can be utilized to efficiently solve the problem~\cite{keller2014response,liakopoulos2019no,chouayakh2022towards,chen2017online,ali2012adaptive,jiang2013optimal}. Control theory and queuing theory are also common theoretical tools for designing resource management solutions~\cite{gias2019atom,ali2012adaptive,grimaldi2015feedback,baarzi2021showar}. Despite the favorable theoretical characteristics of such solutions, real-life cloud applications generally fail to satisfy the desired problem structure due to varying workload profiles and other cloud uncertainties~\cite{alipourfard2017cherrypick}.

\noindent \textbf{Predictive approaches using machine learning (ML).} To mitigate over-provisioning overhead and human effort of heuristic-based solutions, predictive approaches predict future workload or system behavior with past statistics and adjust resource allocation in advance to meet future application needs. This type of approaches usually employs well-established machine learning models, such as linear regression~\cite{bodik2009statistical,venkataraman2016ernest}, support vector machine~\cite{yadwadkar2014wrangler} and various types of neural networks~\cite{hazelwood2018applied,zhang2019mark,gan2019seer,park2021graf,luo2022power,wang2022deepscaling,borkowski2016predicting}. Although effective in certain conditions, these ML-based approaches have their intrinsic limitations. First, to deploy such solutions, an exclusive profiling/training phase is generally needed, which can be costly and not available in production-level realistic systems. Second, such ML-based solutions perform best with general workloads or workloads with repeating patterns similar to their training data, but they adapt poorly to fluctuating workloads~\cite{zhang2019mark}. Moreover, training data quantity and quality impact the performance of an ML model to a significant extent. It is also non-trivial and requires specialized domain knowledge to select representative training data and costly retraining is often needed if workload shift happens.


More recently, Reinforcement learning (RL) has captured attention from the resource management community~\cite{qiu2020firm,mao2019learning,mao2019park,rossi2019horizontal}, thanks to its ability to interact with the environment while optimizing its resource allocation actions. However, apart from the fact that RL frameworks also need to pre-train their agents and hence share similar limitations to the aforementioned ML models, they usually fail to achieve a convergence guarantee. Also, in RL models, actions taken are in turn affecting the environment (i.e., the states), while in real-life clouds, many environment variables are independent of actions, such as workload uncertainty which comes directly from the end users.

\subsection{Bandit Algorithms}\label{sec:2.2}
The limitations of existing work suggest that an ideal resource orchestration framework should optimize resource allocation decisions in an online manner with minimum model pre-training and human intervention. More importantly, it should work efficiently in today's complex containerized cloud, taking various cloud uncertainties into account and fitting in different cloud settings. To this end, we resort to the contextual bandit approach~\cite{krause2011contextual}, a data-efficient non-parametric solution. Contextual bandit is an extension of the well-studied Multi-Armed Bandit (MAB) problem~\cite{slivkins2019introduction} by incorporating contextual information about uncontrollable environment variables, such as cloud uncertainties in the cloud computing context. The original MAB problem is a sequential optimization problem, where a player sequentially selects from a finite set of options with probabilistic rewards to maximize the total reward over time. Bayesian Optimization (BO) is a continuous variant of the MAB problem which aims to find the optimizer of a black-box function by incrementally building a model of the objective function. Although part of our control domain (e.g., fine-grained container resource scaling) can be considered continuous which makes our problem essentially a BO with contextual extension, we stick to the term contextual bandits throughout this paper to highlight the contextual nature and align with the theoretical literature.

\noindent \textbf{Bandit algorithms in computer systems research.} Due to the ability to model arbitrary performance functions, bandit algorithms have also been employed in computer system-related research, such as database parameter tuning~\cite{cereda2021cgptuner,zhao2022dremel,marcus2017releasing} and VM configuration selection~\cite{alipourfard2017cherrypick,liu2019accordia,hsu2018arrow}. Dremel and CGPTuner~\cite{cereda2021cgptuner,zhao2022dremel} use bandit algorithms to fine-tune DBMS-specific parameters and the sole objective is to maximize database performance without constraints, while we focus on a lower level of resource orchestration and consider the performance-cost tradeoff. The closest works to ours are Cherrypick~\cite{alipourfard2017cherrypick} and Accordia~\cite{liu2019accordia}. Cherrypick is among the first works to apply bandit algorithms to systems research, aiming to pick the best VM configuration using Bayesian Optimization for big data analytical jobs. It uses Expected Improvement (EI) as its acquisition function, which lacks a convergence guarantee. Accordia studies the exact same problem, and advances one step further by employing the recent GP-UCB algorithm~\cite{srinivas2009gaussian} with convergence guarantee. However, both Cherrypick and Accordia have inherent limitations which prevent them from being readily applicable to the current containerized cloud. First, both works study the VM configuration selection problem where only a finite set of options are available, while finer-grained, almost-continuous control is possible for containers, as mentioned in Section~\ref{intro}. Second, both Cherrypick and Accordia focus on \emph{recurring} analytical jobs, whose workload patterns are regular and predictable. Therefore, they are implicitly using the first few runs of the recurring job as the training phase and thus cannot generalize to workload variations. Last but not least, their performance objectives are solely dependent on the actions taken, and they assume infinite resources without considering the uncontrollable cloud uncertainties and resource-limited private clouds. {\myName}, on the other hand, is uncertainty-aware and generalizes to different cloud workloads and settings. We would also like to mention RAMBO~\cite{li2021rambo}, a BO-based resource allocation framework for microservices. Although RAMBO solves a similar problem to our work, technical details of implementation and design choices are not sufficiently provided in the paper. A detailed comparison between {\myName} and closely related works is summarized in Table~\ref{table:comparison}. 

\section{Problem Analysis}\label{obs}
In this section, we show through experimental analysis 
important observations which motivate our work. To justify the complex performance-cost relationship and the substantial impact of cloud uncertainties on application performance in a containerized cloud, we set up a cloud-hosted testbed consisting of 16 VMs (see Section~\ref{exp} for detailed specifications) to run benchmarking jobs. All jobs are submitted as Kubernetes-managed containers unless otherwise specified. 

\begin{figure}[t]
  \begin{subfigure}{\linewidth}
        \includegraphics[width=\linewidth]{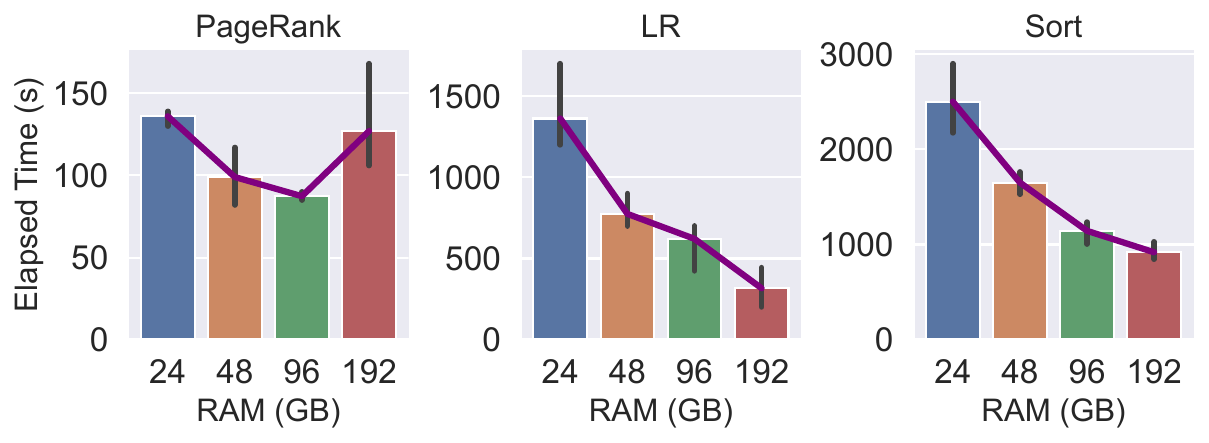}
        \caption{Spark on Kubernetes.}\label{fig:spark_k8s}
  \end{subfigure}
  \begin{subfigure}{\linewidth}
        \includegraphics[width=\linewidth]{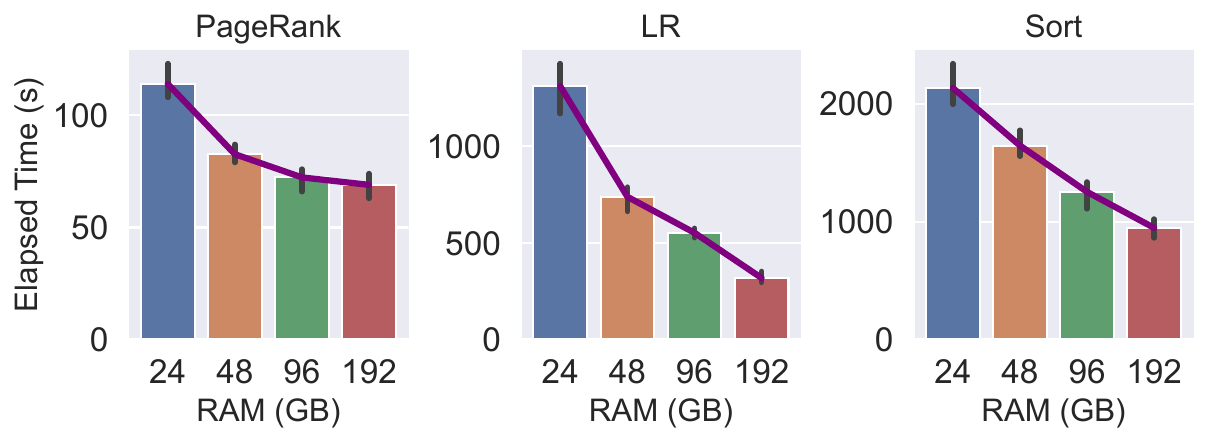}
        \caption{Spark on VMs.}\label{fig:spark_vm}
  \end{subfigure}
  \caption{Performance of representative Spark analytical workloads under different RAM allocations.\label{fig:spark_performance}}
\end{figure}

\begin{figure*}[t]

\begin{minipage}[b][1\totalheight][b]{0.24\textwidth}%
\begin{center}
\includegraphics[width=1\columnwidth]{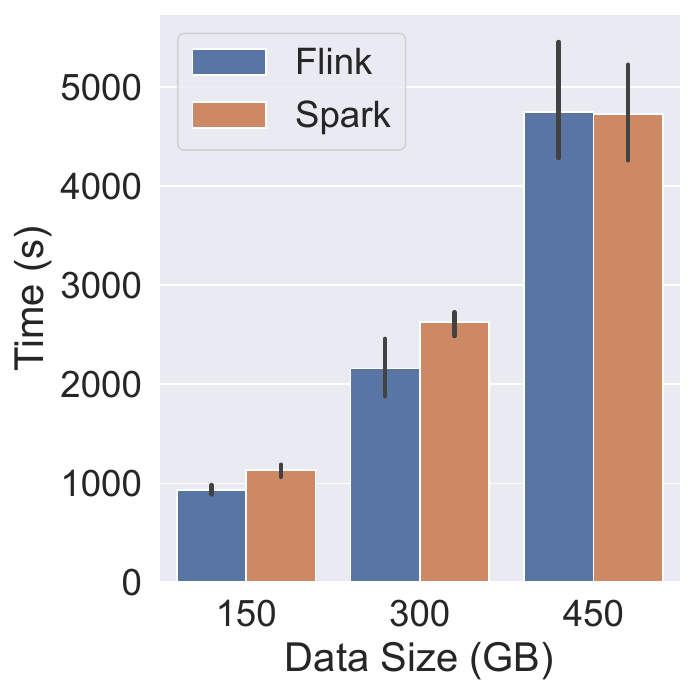}
\captionof{figure}{Sort performance on Flink and Spark.\label{fig:sort}}
\par\end{center}
\end{minipage}
\begin{minipage}[b][1\totalheight][b]{0.48\textwidth}%
\begin{center}
\includegraphics[width=1\columnwidth]{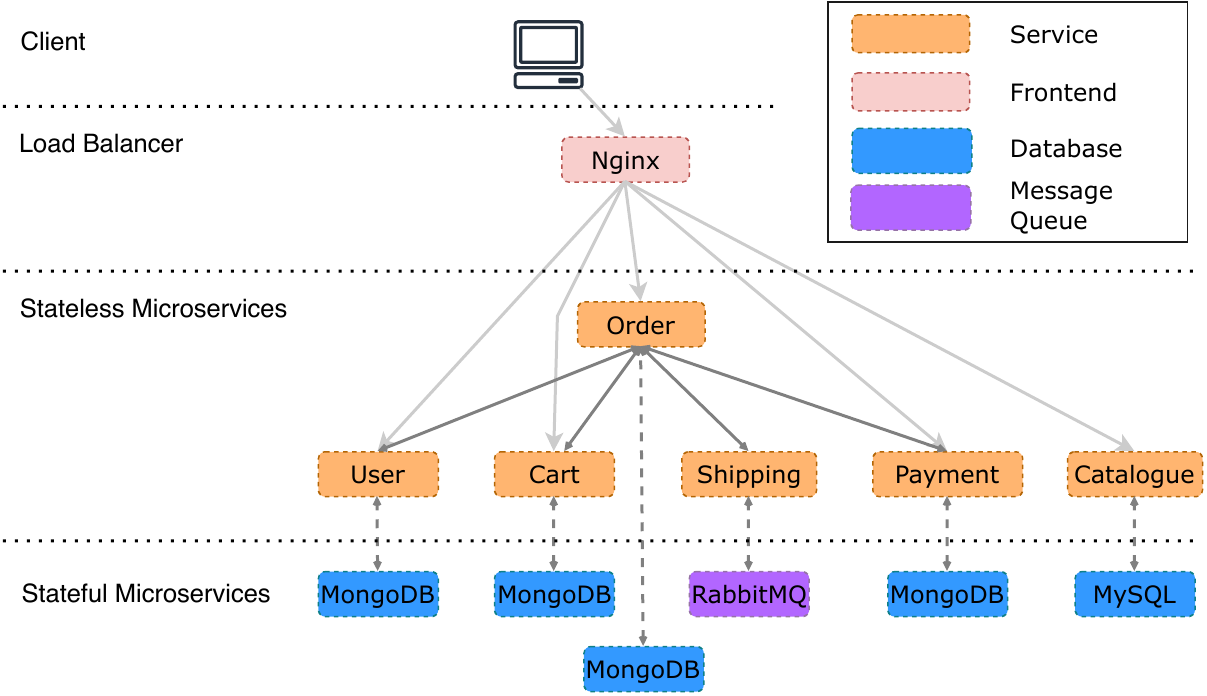}
\captionof{figure}{The \texttt{Sockshop} architecture.}\label{fig:sockshop}
\par\end{center}
\end{minipage}%
\begin{minipage}[b][1\totalheight][b]{0.24\textwidth}%
\begin{center}
\includegraphics[width=1\columnwidth]{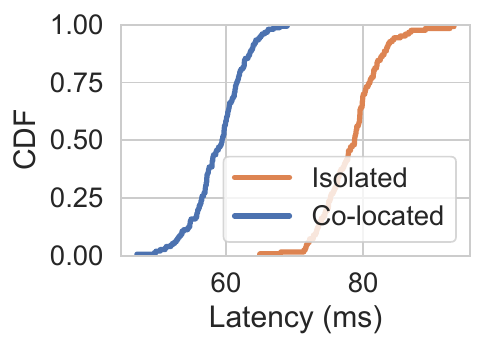}
\captionof{figure}{CDF of microservice end-to-end latency with two node affinity rules.}\label{fig:ms_performance}
\par\end{center}
\end{minipage}
\end{figure*}

\noindent \textbf{Non-structural performance-cost relationship.} To study the relationship between application performance and allocated resources, we benchmark three representative analytical workloads running on the native Spark Operator on Kubernetes~\cite{spark_on_k8s}: PageRank, Sort and Logistic Regression (LR). PageRank is a graph-processing algorithm for which we use the Pokec social network graph data~\cite{takac2012data} with 1.6M vertices and 30M edges. We use \texttt{gensort}~\cite{cloudsort} to generate 150GB of randomly permuted 100-byte records for sorting. For LR, we use a 4-year span of \textasciitilde 400k stock price records from the Nifty 100 Index~\cite{nifty} to train the model. Experiments are repeated five times and the results are shown in Figure.~\ref{fig:spark_performance}(a). While allocating more RAM generally leads to better performance, beneficial theoretical attributes such as linearity and convexity are not manifested in this relationship. For example, LR does not suffer from performance gain saturation when given excessively more RAM, displaying an over 2x performance improvement with increasing RAM allocation from 96GB to 192GB because it benefits from more RAM as a memory-bound job. More interestingly, the performance-cost relationship can even be non-monotonic, meaning more resources does not necessarily lead to performance improvement, as can be observed for PageRank. This is largely due to the fact that PageRank is an iterative network-intensive algorithm where data shuffling between not-co-located containers is needed in each operation. In this case network bandwidth is the major bottleneck instead of RAM.

We repeat the same experiments using identical configurations on the vanilla Spark cluster deployment without involving containers and report the results in Figure.~\ref{fig:spark_performance}(b). Although the performance metrics and the performance-cost relationship patterns are similar to the containerized setting, an important finding is that the variance of performance measurements in the VM-based setting (indicated by black confidence intervals on each bar) is much smaller. The stability is in part owing to the more mature architectural support, but also corroborates our insight that greater uncertainties and anomalies are introduced in a containerized cloud. In fact, we do observe more frequent Spark executor errors and restarts on Kubernetes. 

\noindent \textbf{Impact of cloud uncertainties.} We also show that besides workload intensity, other uncontrollable cloud uncertainties can also significantly impact application performance. To better model adverse situations in a shared cloud, we apply interference injection across experiments to create random resource contention ~\cite{qiu2020firm}, including CPU utilization, RAM bandwidth and network latency and bandwidth. The interferences’ occurrence follows a poisson process with average rate of 0.5 per second. The intensity of each interference is uniformly and independently chosen at random between [0,50\%] of the total capacity. We first study the performance of sorting varying sizes of data on the Kubernetes deployment of Spark and Flink~\cite{k8s_on_flink}. All experiments are conducted five times with the same resource configuration (36 CPU cores and 192GB of total RAM) and identical data for one size. The results are shown in Figure~\ref{fig:sort}. We can observe that the variance across multiple runs increases with data size, reporting a coefficient of variation of up to 23\% for Spark and 27\% for Flink, indicating that application performance can be quite variable due to cloud uncertainties other than workload, especially with a large volume of data which is common in the current ``big data era''. From the performance discrepancy between Spark and Flink, we can also see that the performance is platform-dependent, meaning that even if we have found the optimal resource configuration for one specific workload, it is not readily transferable to other platforms running the same workload and thus additional configuration tuning may be required.

The impact of cloud uncertainties can be even more serious for microservice applications, due to their complicated calling graphs and the resulting inter-container communication patterns~\cite{luo2021characterizing,zhou2018fault}. Towards this end, we deploy an example microservice application \texttt{Sockshop}~\cite{sockshop} consisting of 10+ stateless and database microservices which simulates an online shopping web service. The architecture of \texttt{Sockshop} is shown in Figure~\ref{fig:sockshop}. It is evident that the \texttt{Order} microservice can be a performance bottleneck due to its connection with several other microservices. With the same resource configuration and workload, we compare the end-to-end latency of two affinity rules and show the Cumulative Distribution Function (CDF) in Figure~\ref{fig:ms_performance}. We can find that if we forcefully isolate  \texttt{Order}  from other microservices (by setting node-affinity rules for corresponding pods in Kubernetes), the performance is 26\% worse in terms of P90 latency than the case where we try to colocate \texttt{Order} with other microservices in a best-effort manner. This finding further verifies our claim that the impact of non-workload uncertainties can be significant, and amount-irrelevant resource orchestration decisions can also be deciding factors for application performance.

\section{{\myName} Design}
\label{design}

In this section, we present {\myName}, our \underline{d}ynamic \underline{r}esource \underline{o}rchestrator for the co\underline{n}tain\underline{e}rized cloud. Starting with a brief introduction of contextual bandits and why it is a promising choice for the problem context, we then detail our design and algorithms under both public and private cloud settings. Finally, the implementation and domain-specific optimizations are discussed which complement practically our algorithmic contribution.

\subsection{Overview of Contextual Bandits}\label{sec:4.1}


As briefly discussed in Sec.~\ref{sec:2.2}, it is natural to deduce the mapping from contextual bandits to the cloud resource orchestration problem. The ultimate goal is to dynamically adjust resource allocation decisions to optimize an objective value (e.g., performance and/or cost) in the presence of time-variant cloud uncertainties. Formally speaking, we want to find the best resource configuration $x^*$ from action space $\mathcal{X}$ with uncertainty context $\mathbf{\omega}\in \Omega$ such that the objective function $f$ is optimized:

\begin{equation}\label{eq:f}
    x^* = \argmax_{x\in \mathcal{X}} f(x,\mathbf{\omega})
\end{equation}

From this formulation, we can see that $f$ is dependent on not only the decision variable $x$, but also the context $\mathbf{\omega}$. The output of $f$ can be any scalar value that is of the most interest to the user. Common choices include application performance indicators (e.g., latency, throughput, response time), utility, and cost. Note that~(\ref{eq:f}) is also often formulated as a minimization problem if $f$ is a cost function or captures latency/response time, but the essence of the problem remains unchanged. The action $x$ and context $\mathbf{\omega}$ are vectors with domain-specific dimensions, containing all possible resource orchestration decisions and contextual parameters, respectively. We discuss the concrete dimensions we consider in our problem context in Sec.~\ref{setup}. Since the objective function has no structural relationship with the resource orchestration actions, as we point out in Sec. 3, we can only obtain an objective value by querying the corresponding action. In this case, an exhaustive search of the optimal action is clearly intractable, especially when the action space $\mathcal{X}\in \mathrm{R}^d$ is a continuous domain, and the dimension $d$ is high. 

Towards this end, the contextual bandit approach significantly reduces the search cost by intelligently guiding the next action to search for in an iterative optimization process. Specifically, in each time step $t$, the optimization agent receives a context $\mathbf{\omega}_t$ from the environment. Based on the context, the agent then chooses an action $x_t$ from the action space $\mathcal{X}$, executes this action, and then receives a reward $y_t = f(x_t, \mathbf{\omega}_t) + \varepsilon_t$ as a result of the action taken, where $\varepsilon_t$ is a Gaussian noise $\varepsilon_t \sim \mathcal{N}(0,\sigma^2)$. The noise term well encapsulates the fact that in practice we can only observe a perturbed function value due to unavoidable measurement error. The optimization process then proceeds on to time step $t+1$ with the reward-input pair $(y_t, x_t,\mathbf{\omega}_t)$ appended to the history information to further guide searching in the next iteration.

To evaluate the quality of the actions taken, we use \emph{cumulative regret} $R_T$ which measures the cumulative performance gap over the complete algorithm running span of $T$ time steps, a common metric to assess an online sequential optimization algorithm~\cite{srinivas2009gaussian}:
\begin{equation}
    R_T = \sum_{t=1}^T \left(\max _{x^{*} \in \mathcal{X}} f\left( x^{*}, \mathbf{\omega}_t\right)-f\left(x_t, \mathbf{\omega}_t \right)\right)
\end{equation}
A desired property for an efficient online algorithm is to have \emph{sub-linear regret growth}: $\lim_{T\rightarrow \infty}{R_T / T \rightarrow 0}$, meaning that we can quickly find (near-)optimal actions so that the performance gap converges to zero relatively fast. As we will show in the following sections, {\myName} achieves sub-linear regret growth in both public and private cloud settings.

\subsection{Public Cloud: Cost-aware Performance Optimization}
We first propose our contextual bandit-based algorithm to jointly optimize application performance and resource cost in public cloud environments where computational resources are unlimited.

\noindent \textbf{Why can we assume infinite resources?} It seems natural to assume that computational resources are infinite on public clouds, as previous works~\cite{alipourfard2017cherrypick,liu2019accordia} also instinctively did. While the assumption is plausible given the massive scale of major cloud providers\footnote{For example, it was reported that over 1.4 million servers were hosted on AWS back in 2014~\cite{aws_servers}.}, it may not be readily justifiable from the perspective of individual users or small businesses. For instance, if users are at the edge of their budget for cloud resource renting, they may not be willing to acquire more resources even if that can bring better application performance. In fact, this assumption can be rationalized by cost-saving incentives provided by public cloud, such as Spot Instance~\cite{aws_spot} and Burstable Instance~\cite{aws_burstable} on AWS\footnote{Similar offerings are also available on other major providers, such as Spot Virtual Machines on Azure~\cite{azure_spot} and Google Cloud~\cite{gcp_spot}.}. Spot instances are preemptive, low-priority VMs at a considerably lower price than on-demand instances\footnote{Here, we focus on the cost-saving benefit of spot instances. Users may choose to only deploy non-critical jobs and stateless microservices on spot instances to minimize issues due to potential revocations.}. Burstable instances are VMs that have the ability to ``burst'' to a significantly higher resource configuration than their normal capacity to handle ephemeral and infrequent workload peaks, which is much cheaper than on-demand instances with the same full capacity. We profile the cost-saving effects of spot and burstable instances by issuing the same batch processing (Sort) and microservice workload with the regular instance \texttt{m5.large} as baseline. Table~\ref{tab:incentive} depicts the normalized cost savings of running the same workload across cloud incentive combinations. We observe an up to 7.19x cost saving by employing burstable spot instances and 6.1x cost savings with spot instances alone, both showing notable cost efficiency over on-demand regular instances. Therefore, by judiciously adopting these cloud incentives, one can expect a significant cost reduction achieving the same application performance, meaning that under the same budget, a significantly larger resource configuration search space is available, and this in turn justifies the infinite resources assumption.

\begin{table}[t]
\begin{tabular}{l|l|l|l|}
\cline{2-4}
 & \texttt{m5.large} & Spot only & Spot + Burstable \\ \hline
\multicolumn{1}{|l|}{Batch jobs} & 1x & 6.10x & 7.19x \\ \hline
\multicolumn{1}{|l|}{Microservices} & 1x & 5.28x & 6.73x \\ \hline
\end{tabular}
\caption{Normalized cost savings from cloud incentives.}
\label{tab:incentive}
\end{table}

\begin{figure}[t]
  \centering
  \includegraphics[width=\linewidth]{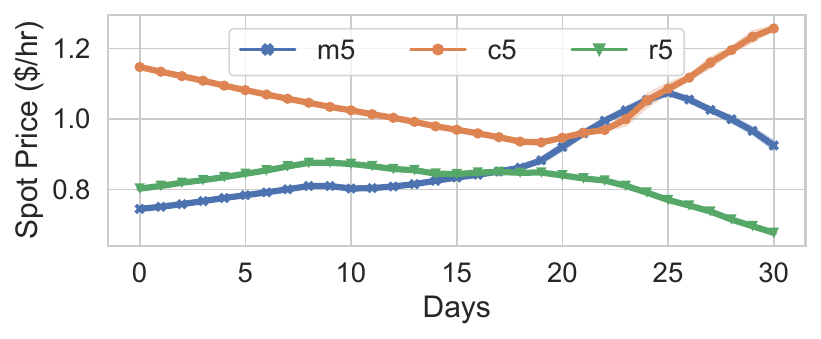}
  \caption{Spot instance prices from April 2023 for m5.16xlarge, c5.18xlarge and r5.16xlarge instance types on AWS.}\label{fig:spotprice}
\end{figure}

Another interesting finding is that spot prices can vary drastically with time in an unpredictable manner. Figure~\ref{fig:spotprice} shows spot prices of three instance types over a 1-month time span, which exhibit no regular patterns and vary across instance types to a great extent. This suggests that the spot price is an additional contextual dimension to be considered which can greatly impact the resource cost.

\noindent \textbf{Problem formulation.} Given the assumption of unlimited resources, the optimization objective on public clouds is to keep a balance between application performance and the monetary resource cost. Formally, our optimization problem can be formulated as maximizing a reward function $f$:
\begin{align}
    \max_{\mathbf{x_t}}\quad &f(x_t,\mathbf{\omega}_t) = \alpha p(x_t,\mathbf{\omega}_t) - \beta c(x_t,\mathbf{\omega}_t)\\
    s.t.\quad &\mathbf{x_t}\in\mathcal{X}, \mathbf{\omega}_t\in\Omega, \quad\forall{t}
\end{align}
where $p(x_t,\mathbf{\omega}_t)$ is the application performance indicator which can be measured at the end of each time step $t$; $c(x_t,\mathbf{\omega}_t)$ is the resource cost associated with the resource orchestration decision $x_t$ and cloud uncertainties enclosed in the context $\mathbf{\omega}_t$. $\alpha$ and $\beta$ are configurable weights that capture a user's preference between performance and cost.

\noindent \textbf{How to guide the search process?} By a sequential optimization process, contextual bandit-based algorithms are able to learn more about the objective function $f$ in every iteration with newly observed data resulting from evaluating $f$ at point $(x_t,\omega_t)$. Therefore, a key design choice of contextual bandits algorithms is to determine how to choose the next point to evaluate so as to learn the most about the objective function. Towards this purpose, we first need to put a surrogate model on $f$ so that it can be efficiently updated iteratively. In {\myName}, we choose Gaussian Process (GP)~\cite{williams2006gaussian}, a common choice adopted by prior works~\cite{alipourfard2017cherrypick,liu2019accordia,cereda2021cgptuner,li2021rambo}. As a non-parametric model, GP assumes a function is sampled from a Gaussian distribution, which adds a minimal smoothness assumption on the objective function such that function values evaluated at close inputs will also be close. Formally, let $z\in \mathcal{X} \times\Omega$ be a joint action-context pair, a GP$(\mu, k)$ is fully specified by its mean function $\mu(z)=\mathbb{E}[f(z)]$ and covariance or kernel function $k(z,z') = \mathbb{E}[f(z)-\mu(z)(f(z')-\mu(z'))]$ which acts as the data-independent prior distribution. Now define that $y_t = f(z_t) + \epsilon_t$ is a noisy sample of true function value $f(z_t)$ and $\mathbf{y}_T = [y_1,y_2,\cdots,y_T]$ are evaluated at points $Z_T = [z_1,z_2,\cdots,z_T]$ (the past data points). Assume now we are given a new $z^*$ and would like to infer the function value $f^*$, we can get a closed-form posterior distribution which is also a GP with the following mean and variance:

\begin{align}
    \mu_T\left(z^*\right)&=\mathbf{k}_T\left(z^*\right)^\top\left(\mathbf{K}_T+\sigma^2 \mathbf{I}\right)^{-1} \mathbf{y}_T\label{posterior1} \\
\sigma_T^2\left(z^*\right)&=k\left(z^*, z^*\right)-\mathbf{k}_T\left(z^*\right)^\top\left(\mathbf{K}_T+\sigma^2 \mathbf{I}\right)^{-1} \mathbf{k}_T\left(z^*\right)\label{posterior2}
\end{align}

where $\mathbf{k}_T(z^*) = [k(z_1,z^*),k(z_2,z^*),\cdots,k(z_T,z^*)]$ and $\left[\mathbf{K}_T\right]_{i j}=k\left(z_i, z_j\right)$ is the kernel matrix. We choose the widely adopted Matern kernel with $\nu = \frac{3}{2}$ following empirical practices. These analytical equations allow us to efficiently infer the function value at a new point based on previous observations and action-context pairs.

Another key element of contextual bandits is to determine how to suggest the next point to evaluate so as to learn most about the objective function. This is achieved by choosing the point maximizing the \emph{acquisition function}, a function that assesses the quality of an action point and is much cheaper to optimize than the original objective function. Next to other popular choices such as Probability Improvement (PI), Expected Improvement (EI) and Thompson Sampling (TS)~\cite{vermorel2005multi, chowdhury2017kernelized}, we choose Upper Confidence Bound (UCB)~\cite{srinivas2009gaussian}, the update rule of which is given as follows:
\begin{equation}
    x_t = \argmax_{x\in\mathcal{X}}\;\mu_{t-1}(x, \omega_t) + \sqrt{\zeta_t}\sigma_{t-1}(x,\omega_t)\label{eq:UCB}
\end{equation}
An important rationale behind choosing UCB, as can be perceived from the equation, is that it efficiently balances \emph{exploration} of undiscovered resource configurations and \emph{exploitation} of configurations that have already been observed to be well-performing. The hyperparameter $\zeta_t$ serves to balance the tradeoff: choosing a small $\zeta_t$ indicates we value the first mean term more, hence will more likely select an action close to one which previously led to better performance; choosing a large $\zeta_t$, on the other hand, focuses more on the variance term so that under-explored actions with higher uncertainty are more likely to be selected. Moreover, in the GP setting, UCB is superior in terms of both computational efficiency~\cite{wilson2020efficiently, vakili2021scalable} and convergence rate~\cite{chowdhury2017kernelized} compared to alternatives such as GP-TS.

Combining these design choices, we summarize our GP-UCB-based online resource orchestration algorithm in Algorithm~\ref{alg1}. 

\begin{algorithm}
\caption{Contextual Bandits for Public Clouds\label{alg1}}
\begin{flushleft}
\textbf{Input:} Performance-cost balance weights $\alpha, \beta$;\\
\hspace*{0.92cm}Action Space $\mathcal{X}$; \\
\end{flushleft}
\begin{algorithmic}[1]
  \State $S_0 \leftarrow \emptyset$; \Comment{$S_t$ stores action-context pairs up to time $t$}
  \State $\mathbf{y}_0 \leftarrow \emptyset$; \Comment{$\mathbf{y}_t$ stores noisy rewards up to time $t$}
  \For{$t = 1,2,\cdots$}
  \State Observe current context $\omega_t$;
  \State Select resource configuration $x_t$ according to~(\ref{eq:UCB});
  \State Observe noisy reward $y_t = f(x_t,\omega_t) + \epsilon_t$;
  \State $S_t \leftarrow S_{t-1}\cup(x_t,\omega_t)$;
  \State $\mathbf{y}_t \leftarrow \mathbf{y}_{t-1}\cup y_t$;
  \State Update $\mu_{t}$ and $\sigma_{t}$ by posterior update rule~(\ref{posterior1})-(\ref{posterior2});
  \EndFor
\end{algorithmic}
\end{algorithm}

\noindent \textbf{Regret analysis.} A desired property of a bandit algorithm is to have sub-linear cumulative regret growth. Our algorithm achieves this with high probability by setting appropriate hyperparameters, as shown in the following theorem:

\begin{theorem}
Let $\delta\in (0,1)$, $\forall T\geq1$, the cumulative regret of Alg.~\ref{alg1} is upper bounded by $O(\sqrt{T\gamma_T\zeta_T})$ with high probability. Precisely,
\begin{equation}
    Pr\{R_T\leq \sqrt{C_1T\gamma_T\zeta_T} + 2\} \geq 1-\delta
\end{equation}
where $C_1 = \frac{8}{\log (1+\sigma^{-2})}$, $\zeta_t = 2B^2 + 300\gamma_t\log^3 (\frac{t}{\delta})$.
\end{theorem}

Here, $\gamma_T$ is the maximum information gain in the order of $O(T^l\log T)$ where $l<1$. $B\geq ||f||_k$ is the upper bound of the Reproductive Kernel Hilbert Space (RKHS) norm of $f$, a common assumption in bandit algorithms. Due to space constraint, please refer to~\cite{thm_proof} for proof of the theorems.

\subsection{{Private cloud: Resource-constrained Performance Optimization }}
For security or data privacy concerns, organizations often resort to a private cloud solution instead of running their jobs on a public cloud. A private cloud is a self-hosted computing cluster that the organization has full control of. The organization is also able to fully unlock the power of computing nodes by customizing their hardware and software architectures tailored to its own needs which is often constrained on public clouds. Compared to the pay-as-you-go model on public clouds, organizations pay the resource cost upfront at the purchase of the hardware to build the private cloud. The update cycle is generally up to several years when the hardware is too old or the business scale has been significantly expanded. In this case, any resource orchestration decision must respect the private cloud's total resource limit, which is a hard constraint. The optimization objective under such scenarios is thus optimizing application performance subject to the hard resource constraints~\cite{sui2015safe,sui2018stagewise,amani2020regret}. Formally, the resource orchestration optimization problem in the private cloud can be formulated as:
\begin{align}
    \max_{\mathbf{x_t}}\quad &p(x_t,\mathbf{\omega}_t)\\
    s.t.\quad &x_t\in\mathcal{X}_t^S, \mathbf{\omega}_t\in\Omega, \quad\forall{t}
\end{align}
where $\mathcal{X}_t^S$ is the \emph{safe} set from the action domain at time step $t$ so that actions can only be selected from the safe set to comply with resource constraints. Specifically, denote $P_{max}$ as the resource constraint and let $P(x_t,\omega_t)$ be the total resource usage resulting from action $x_t$ and context $\omega_t$ at time step $t$, we have
\begin{equation}
    \mathcal{X}_t^S = \{x_t\in\mathcal{X}:P(x_t,\omega_t) \leq P_{\max}\}
\end{equation}

Note that $P(x_t,\omega_t)$ and $P_{\max}$ contain multiple dimensions in practice. Each of the dimensions is a resource type (e.g., CPU, RAM, and network bandwidth) and has its own limit in a private cloud. For presentation brevity, here, we abstract them as an overall constraint, without loss of generality. Moreover, $P(x_t,\omega_t)$ is also an unknown function since it depends on the contextual variables $\omega_t$ as well. This is reasonable since resource contention is common in a shared cloud (a private cloud can also be shared within the organization across several development teams). As a result, the performance indicator function $p(x_t,\mathbf{\omega}_t)$ and the resource usage function $P(x_t,\omega_t)$ need to be modelled separately. At each time step $t$ throughout the optimization process, our algorithm needs to select an action $x_t$ from the safe set ${X}_t^S$ so that the performance $p(x_t,\omega_t)$ is optimized. Towards this end, we use two GPs to model the performance function and the resource function, respectively. Reusing the notation $z\in \mathcal{X}^S \times \Omega$ as a joint safe action-context pair, at each time step $t$, noisy values of both functions are observed as $y_t = p(z_t) + \epsilon_t$ and $\phi_t = P(z_t) + \epsilon_t$. We now present our solution in Algorithm~\ref{alg2}.

\begin{algorithm}
\caption{Contextual Safe Bandits for Private Clouds}
\label{alg2}  
\begin{flushleft}
\textbf{Input:} Action Space $\mathcal{X}$, Initial Safe Set $\mathcal{X_\zeta^S}$, Exploration Duration $T^\prime$
\end{flushleft}

\begin{algorithmic}[1]
  \State \textbf{Step 1: Exploration}
  \For{$t=1,\cdots,T^\prime$}
  \State Observe context $\omega_t$;
  \State Randomly choose $x_t\in\mathcal{X_\zeta^S}$;
  \State Observe noisy performance $y_t = r(x_t,\omega_t) + \epsilon_t$;
  \State Observe noisy resource usage $\phi_t = P(x_t,\omega_t) + \epsilon_t$;
  \EndFor
  \State \textbf{Step 2: Exploration + Exploitation}
  \For{$t = T^\prime,\cdots$}
  \State Update $\mu_{t-1}$, $\sigma_{t-1}$ by posterior update rule~(\ref{posterior1})-(\ref{posterior2});
  \State Observe context $\omega_t$;
  \State $l_P (x_t,\omega_t) \leftarrow \mu_{P,t-1}(x_t,\omega_t) - \sqrt{\beta_t}\sigma_{P,t-1}(x_t,\omega_t)$;
  \State $u_p (x_t,\omega_t) \leftarrow \mu_{p,t-1}(x_t,\omega_t) + \sqrt{\beta_t}\sigma_{p,t-1}(x_t,\omega_t)$;
  \State $\mathcal{X}_t^S \leftarrow \{x_t\in\mathcal{X}:l_P (x_t,\omega_t)\leq P_{max}\}$; \\\Comment{The new, expanded safe set}
  \State $x_t \leftarrow \argmax_{x\in\mathcal{X}_t^S} u_p(\omega_t,x)$;
  \State Observe noisy performance, resource usage and update corresponding vectors;
  \EndFor
\end{algorithmic}
\end{algorithm}

The core idea of this algorithm is a two-phase process. In the first phase, starting from a guaranteed safe set, the algorithm is dedicated to exploration by randomly querying actions to gather more information to characterize the safe set. The second phase acts similarly to Alg.~\ref{alg1} which balances exploration and exploitation by following the GP-UCB procedure to update the posterior GP model. However, on top of the standard GP-UCB algorithm, it leverages information from previous exploration to iteratively expand the safe set based on the lower confidence interval of the resource usage function $P$ (Line 14). We show through the following theorem that Alg.~\ref{alg2} also achieves sub-linear cumulative regret growth:

\begin{theorem}
Let $\delta\in (0,1)$, for sufficiently large $T\geq1$, the cumulative regret of Alg.~\ref{alg2} is upper bounded by $O(\sqrt{T\gamma_T\zeta_T})$ with high probability. Precisely,
\begin{equation}
    Pr\{R_T\leq BT' + \sqrt{C_1T\gamma_T\zeta_T}\} \geq 1-\delta
\end{equation}
where the parameters $C_1, \gamma_T$ and $\zeta_T$ take the same value as the previous theorem.
\end{theorem}

\subsection{{\myName} Implementation}

\begin{figure}[t]
  \centering
  \includegraphics[width=\linewidth]{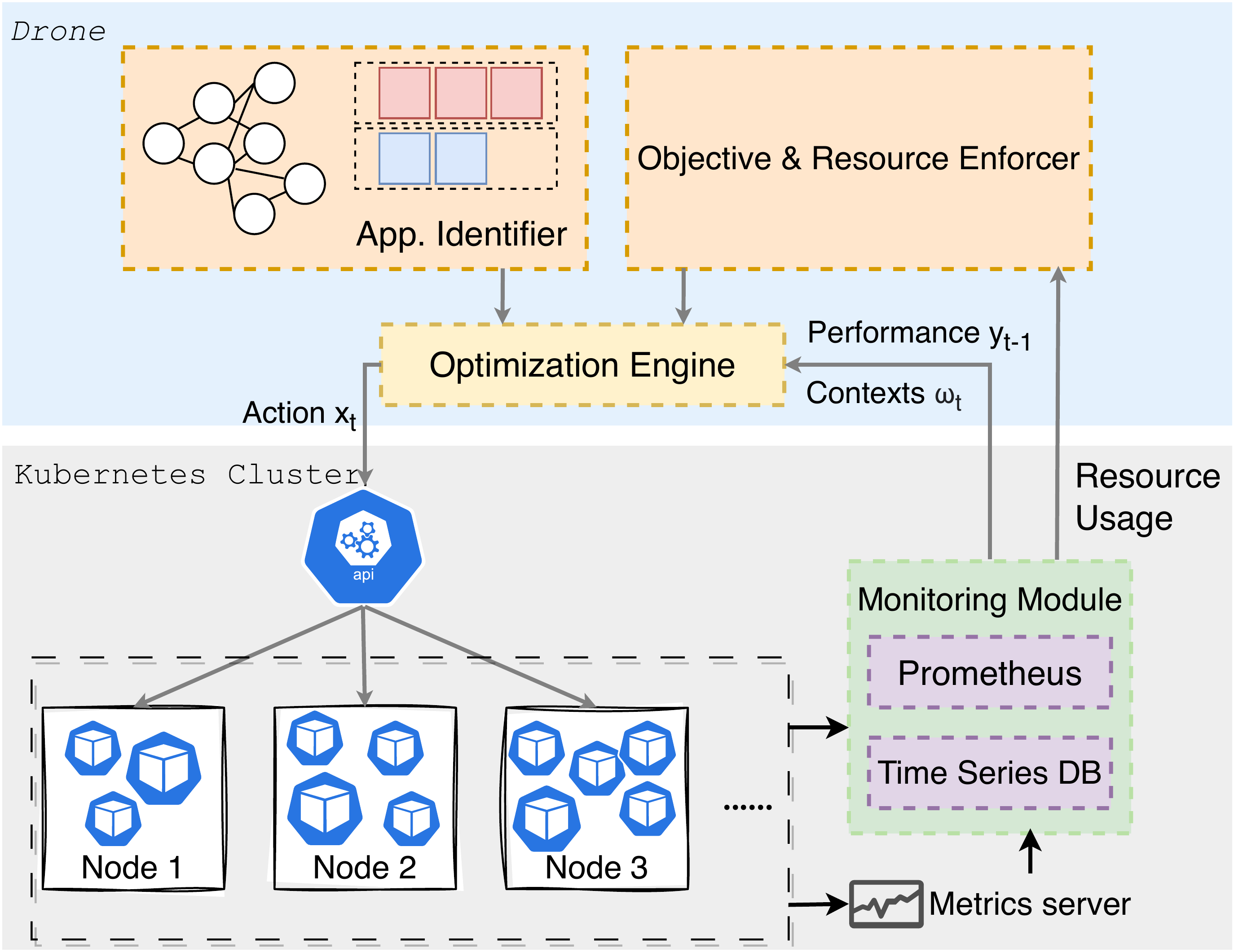}
  \caption{{\myName} Architecture.\label{fig:drone}}
\end{figure}

We implemented a prototype of {\myName} as an integrable resource orchestrator on top of Kubernetes. The overall system architecture of {\myName} is depicted in Figure~\ref{fig:drone}, which contains the following components:

\noindent \textbf{Monitoring Module.} The monitoring module is a key component of the {\myName} framework. It is responsible for periodically collecting both performance metrics and contextual information from the cloud environment. In {\myName}, we choose Prometheus~\cite{Prometheus} for this purpose. Prometheus is a market-leading monitoring system shipping with a time series database and powerful querying capabilities through its specialized query language \texttt{PromQL}. It is able to collect system-level real-time metrics such as CPU, RAM and network bandwidth usage through \texttt{node-exporter}~\cite{node_exporter} along with other potential contextual variables. By exposing a metrics exporter, applications are also enabling Prometheus to collect their performance metrics like throughput and response time. The collected metrics are stored in the time series database which can be efficiently queried upon request. The collected real-time contextual information, along with stored history performance data and action-context pairs, provides input to guide the optimization process of {\myName}'s algorithms.

\noindent \textbf{Application Identifier.} The application identifier helps identify the type of the submitted application to make tailored resource orchestration decisions for batch processing jobs and microservices, respectively. While users can explicitly specify the application type, as discussed in~\ref{sec:opt}, the application identifier is also able to automatically detect the application type if it is evident in the deployment specification. For example, a Spark application has an exclusive \texttt{kind: SparkApplication} specification field which can be easily utilized by the application identifier. 

\noindent \textbf{Objective and Resource Enforcer.} Depending on whether the environment is a public cloud or a private cloud, this module specifies the optimization objective for the optimization engine. Users can tune model parameters here based on their needs, such as performance-cost preference coefficients in the public cloud setting and resource limit in the private cloud setting. In a private cloud, if the user does not specify the desired resource limit, the enforcer will set the limit according to the cluster resource usage. \looseness=-1

\noindent \textbf{Optimization Engine.} As the core part of the framework, the optimization engine is responsible for carrying out the optimization process. Based on the cloud setting set by the application identifier and the enforcer, the optimization engine continuously receives performance and contextual metrics from the monitoring module and suggests a resource orchestration action in each decision period. The action is a combination of container rightsizing and scheduling. Actions are executed by directly interacting with the Kubernetes API server to minimize additional overhead.

\subsection{Cloud-specific Optimizations\label{sec:opt}}

On top of the algorithmic efforts, we also make practical optimizations tailored to the cloud resource orchestration problem context to further improve {\myName}'s usability and efficiency in practice.

\noindent \textbf{Encoding of actions and contexts.} Unlike CPU cores and RAM allocation/usage which take numerical values and thus naturally fit in our contextual bandit-based framework, some action and contextual variables do not readily manifest a numerical representation such as container scheduling decisions from the action space and possible traffic bottleneck from the context space. We address this issue by scalarizing these variables with numerical encoding. For example, we encode the scheduling decisions as a sub-vector $\underline{x} = [x_1,x_2,\cdots, x_m]$ of the entire decision vector $x$, where $m$ is the number of computing nodes or VMs on which a container can be scheduled. The elements $x_i \in \mathbb{N}$ represent the number of containers that should be scheduled to node $i$. Note that having an individual entry for each single node may lead to dimension explosion when the cloud scale is large. However, in practice, we can further group nodes by physical distance into zones within which nodes perceive almost no network latency when communicating with each other. The scheduling decisions will thus be executed at the zone level, significantly reducing dimensionality to the number of zones. This is particularly useful when the cloud is geographically distributed where high latency can be incurred by inter-zone communication. For traffic between nodes, we can use an integer $a\in [0,2^m-1]$ to encode the possible traffic contention, which can be proven trivially by the binomial theorem.

\noindent \textbf{Characterization of applications.} We consider two representative application profiles for {\myName}, namely batch processing jobs and long-running web services in the form of microservices. Also referred to as Best Effort (BE) and Latency Critical (LC) applications in the recent literature~\cite{lu2023understanding,zhang2022workload,cheng2018analyzing}, these two types cover the majority of applications running on a massive-scale production cloud~\cite{lu2023understanding}. When using {\myName} as the resource orchestrator, users have the option to specify their application type as either batch jobs or microservices so as for {\myName} to make specialized decisions. Essentially, {\myName} can run in two modes. For batch jobs, {\myName} follows previous works~\cite{alipourfard2017cherrypick,liu2019accordia} to run the algorithm in a quasi-online manner in order to further leverage the additional optimization opportunities brought forth by recurring workload patterns, but achieves better performance by incorporating other cloud uncertainties and resource constraint compliance under private cloud setting. For microservices, {\myName} runs as a fully online algorithm so that all resource orchestration actions are updated on the fly as time progresses. These options enable {\myName} to make wiser resource orchestration decisions to best fit the application characteristics.

\noindent \textbf{Reducing computational complexity.} A well-known limitation of the GP-UCB approach is the cubic $O(n^3)$ computational complexity associated with the matrix inverse operation, where $n$ is the number of data points. At each time step $t$, all previous data points $(S_t, \mathbf{y}_t)$ are used with the same weight to update the surrogate model, and this gradually becomes computationally intractable as time progresses. Based on the insight that data points from the long-ago past may not be as representative as recent ones in terms of modeling capability, we adopt a sliding-window approach to address this issue. Instead of using them all, we only consider the most recent $N$ data points for updating the model. This significantly reduces the running time complexity of {\myName} over time.  Across our experiments, we set $N=30$ which empirically entails a good balance between algorithm performance and running time. 

\noindent \textbf{Initial point selection.} Just like gradient-based optimization approaches, bandit algorithms usually need a set of initial data samples to bootstrap the function estimation process (e.g., PI- and EI-based approaches). The algorithm performance can depend on the quality of these initial selections to a great degree and designing a selection strategy is non-trivial and domain-specific. Although UCB-based algorithms suffer less from this problem by valuing the variance term to motivate exploration, as also pointed out in~\cite{liu2019accordia}, it is still a practical concern under the private cloud setting since a poorly chosen initial configuration may violate the resource constraints and hence invalidate the algorithm. An intuitive approach is to start from minimum resource configurations which almost surely lie in the safe set and expand possible action choices from there. However, we observe from experiments in Sec.~\ref{obs} that running a job with minimum resources can be problematic. For example, running the same PageRank job on Spark with a total RAM of less than 12GB would leave the job in a halt stage so that no performance metrics can be collected and hence no progress can be made by our algorithm. Therefore, we follow a heuristic approach to allocate half of the currently available resources by querying the monitoring module for the current total resource utilization as a starting point. In our experiments, this proves to be a good selection with a low error rate across workloads. As a safety measure, we also implement a failure recovery mechanism that if a job errors out with no metrics produced in a pre-defined timeout period, it will be restarted with a higher resource configuration at the midpoint of the previous trial and the maximum resources available. \looseness=-1

\section{Experimental Evaluation}\label{exp}

\subsection{Experimental Setup}\label{setup}

\noindent \textbf{Testbed setup.} Our testbed cluster is hosted on Compute Canada~\cite{baldwin2012compute}, a shared national cloud platform. The cluster consists of 16 virtual machines with one control node and 15 worker nodes, a scale comparable to related work. The control node is equipped with 16 vCPU cores and 180GB of RAM, while the worker nodes have 8 vCPU cores and 30GB of RAM. Each node runs Ubuntu 20.04.5 LTS with Linux kernel v5.4. Nodes are interconnected by 10Gb Ethernet in the same data center. Kubernetes v1.25 is deployed as the container orchestration platform.

\noindent \textbf{Applications.} Representative applications for both batch processing jobs and microservices are deployed to evaluate {\myName}. For batch jobs, we benchmark three Spark applications that stress different computational resources, including (1) Spark-Pi, a pi-computation job with configurable number of iterations to control precision as a representative compute-intensive job, (2) PageRank as a jointly memory- and network-intensive job and (3) Logistic Regression to serve as a typical ML training task. For microservices, we use the \textit{Social Network} application containing 36 microservices from DeathStarBench~\cite{gan2019open}, a prevalent microservice banchmarking choice in the research community.  Various calling patterns are present between the microservices to mimic production microservice applications. All microservices are running as Docker containers managed by Kubernetes in the form of pods. 


\noindent \textbf{Workloads.} We use the same workloads as discussed in Sec.~\ref{obs} for batch jobs with configurable intensity. For microservices, we use a 6-hour sample from the Twitter Streaming trace~\cite{twittertrace} which exhibits diurnal patterns, a good representation of real-life web service requests. The workload is generated by \texttt{wrk2}, a multi-threaded workload generator in which several parameters can be fine-tuned to control workload characteristics, such as request rates, distribution and elapsed time. We tailor these parameters to properly fit the workload to the scale of our cluster. Different request types are also customized to randomly stress multiple microservices. Workloads are all sent from a separate node than the ones that host the microservices to simulate external users.

\noindent \textbf{Comparison baselines.} We compare {\myName} with Cherrypick~\cite{alipourfard2017cherrypick} and Accordia~\cite{liu2019accordia} in the case of batch job processing. For the sake of fair comparison, batch jobs are run repeatedly, a case for which Cherrypick and Accordia are specifically designed. For microservices, our baselines include SHOWAR~\cite{baarzi2021showar}, a hybrid autoscaling approach specialized for microservices combining both vertical and control theory-based horizontal autoscaling, and Google Autopilot~\cite{rzadca2020autopilot}. Autopilot is Google's hybrid autoscaler in production use. Its basic model works by a moving-window approach to aggregate resource utilization statistics in recent time windows to produce the resource utilization targets and linearly scale resources with respect to the targets. In both cases, we also run experiments with the default rule-based Kubernetes Horizontal Pod Autoscaler (HPA) and its native scheduler to serve as a standard baseline.

\begin{figure*}[t]
    \centering
    \begin{subfigure}{0.32\textwidth}
        \centering
        \includegraphics[width=\linewidth]{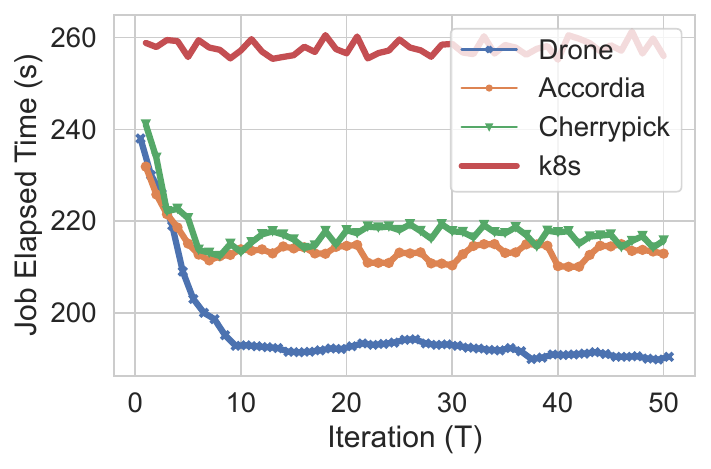}
        \caption{Batch job (LR) elapsed time under the public cloud.}
    \end{subfigure}\hfill
    \begin{subfigure}{0.32\textwidth}
        \centering
        \includegraphics[width=\linewidth]{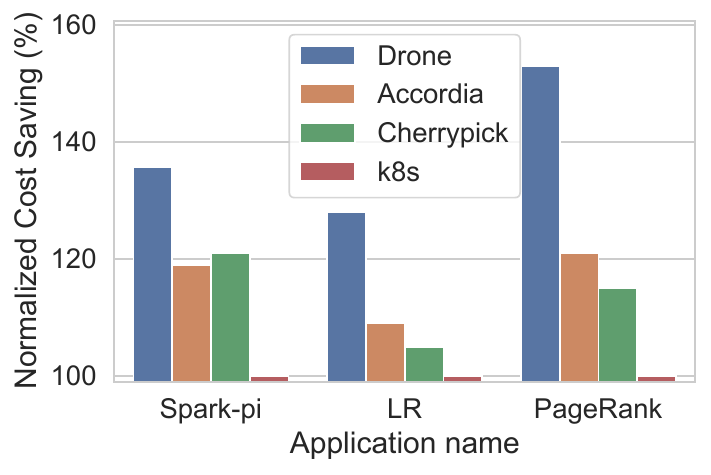}
        \caption{Resource cost savings across batch jobs under the public cloud.}
    \end{subfigure}\hfill
    \begin{subfigure}{0.32\textwidth}
        \centering
        \includegraphics[width=\linewidth]{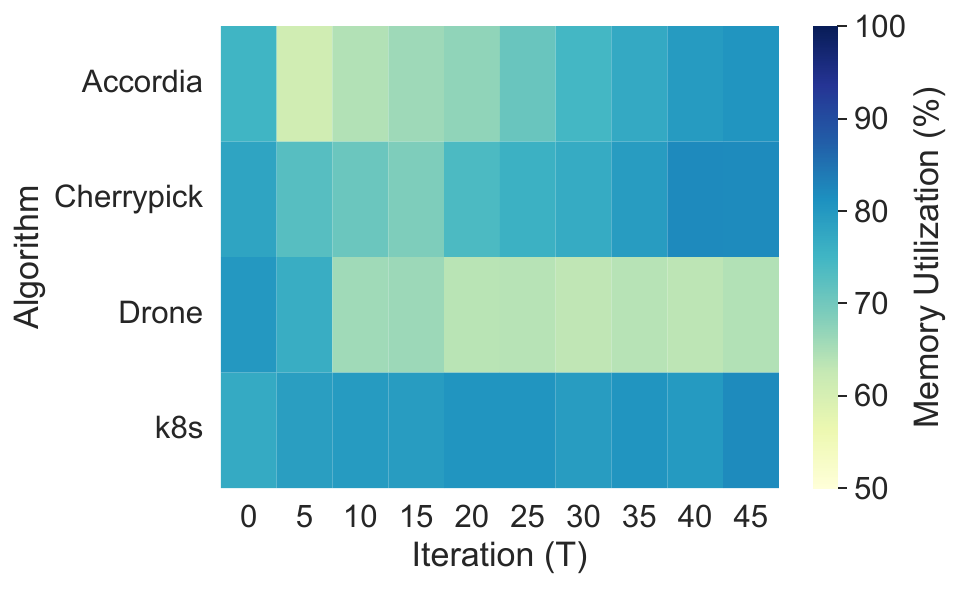}
        \caption{Overall memory utilization for batch jobs under the private cloud.}
    \end{subfigure}
    \caption{Comparison between {\myName} and alternatives for batch processing jobs.\label{fig: eval_batch}}
\end{figure*}

\noindent \textbf{Performance metrics and input spaces.} The performance metrics, utilized as objective functions in our formulations presented in Section~\ref{design}, are measured according to their distinct specifications. The performance indicator $p(x_t,\omega_t)$ represents job elapsed time (i.e., the wall-clock time between its submission and completion) for batch jobs, and P90 end-to-end latency (i.e., the 90th percentile of latency for requests sent) for microservices. To more appropriately model our formulations, the resource cost $c(x_t,\omega_t)$ is estimated using the resource-based pricing scheme adopted by Google cloud~\cite{gcepricing} which charges users by actual resource usage as opposed to AWS which charges users according to VM types. These performance metrics along with low-level resource utilization metrics are collected by the monitoring module every 60 seconds to align with the Prometheus \texttt{scrape-interval}, which is also our decision period when running fully online. In our evaluations, we also group our nodes into four zones with artificial latency between each other using Linux traffic control (\texttt{tc}). Therefore, our action space is a 7-dimension space including the number of pods to schedule to each zone along with CPU, RAM and network bandwidth allocation for each pod. The context space on the other hand contains workload intensity, current CPU, RAM and network bandwidth utilization in the cloud, potential traffic contention and spot prices. To incorporate the influence of spot prices, we randomly fill 10-30\% of the resource cost with spot prices. The spot prices are sampled from the \texttt{E2} instance family spot prices on Google Cloud. In the private cloud setting, the spot price dimension is omitted.

\begin{table}[t]
  \centering
\begin{adjustbox}{width=\linewidth}

\begin{tabular}{l|cc|cc|cc}
\toprule
     & \multicolumn{2}{c|}{Spark-Pi} & \multicolumn{2}{c|}{LR} & \multicolumn{2}{c}{PageRank} \\
\multicolumn{1}{c|}{Framework} & Time(s)  & \# Errors & Time(s)  & \# Errors  & Time(s) & \# Errors\\ \hline
\multicolumn{1}{l|}{k8s} & 53\textpm 2 & 0 & 328\textpm 17 & 1 & 1436\textpm 88 & 4  \\
\multicolumn{1}{l|}{Accordia} & 46\textpm 1 & 0 & 303\textpm 26 & 17 & 1172\textpm 95 & 98  \\
\multicolumn{1}{l|}{Cherrypick} & 43\textpm 1 & 0 & 298\textpm 24 & 13 & 1226\textpm 102 & 107  \\
\multicolumn{1}{l|}{\myName} & 41\textpm 1 & 0 & 226\textpm 9 & 5 & 785\textpm 42 & 9  \\\hline
\bottomrule
\end{tabular}%
\end{adjustbox}
\caption{{\myName} significantly reduces OOM errors by conforming with resource constraints. }
\label{tab:batch_private}%
\end{table}%

\subsection{{\myName} for Batch Processing Jobs}

We start off by investigating {\myName}'s effectiveness in handling recurring batch processing workloads. In the public cloud setting, we set $\alpha = \beta = 0.5$ to model a general user who does not have a specific preference and normalize the performance and cost values to the same magnitude for fair comparison with alternatives. Our results are summarized in Figure.~\ref{fig: eval_batch}. First, Figure.~\ref{fig: eval_batch}(a) depicts the performance measurements for the same LR job running with different schemes in the public cloud setting. Starting from the same starting point, it can be seen that all three bandit algorithm-based approaches are able to improve application performance by learning the performance-input relationship function over time. On the other hand, as a completely reactive rule-based autoscaler, the default Kubernetes solution cannot benefit from history information, and hence only manages to maintain a low performance, which is slightly perturbed over time by environment uncertainties and measurement errors. {\myName} significantly outperforms Cherrypick and Accordia by adaptively learning from the contextual variables while Cherrypick and Accordia are oblivious to such environment changes and can only leverage information from their resource decisions, i.e., the action space. The benefit of considering contextual information can further be observed from the post-convergence behaviour when $T > 10$. Both Cherrypick and Accordia sporadically experience performance oscillations while {\myName} is able to stabilize after convergence. This is because Cherrypick and Accordia regard any performance feedback as the exclusive result of the actions taken. Therefore, whenever a performance discrepancy is observed they will adjust their resource allocations even though it is primarily owing to changes in contextual cloud uncertainties. It is also worth noting that {\myName} converges slightly slower at the 10th iteration compared to the other two schemes which converge around the 7th iteration because the search space is larger in {\myName} due to additional dimensions in the action spaces (e.g., the scheduling vector) and the new contextual dimensions. This is a common performance-dimension tradeoff which we will briefly discuss in Sec.~\ref{sec:discuss}.

\begin{figure*}[t]
    \centering
    \begin{subfigure}{0.32\textwidth}
        \centering
        \includegraphics[width=\linewidth]{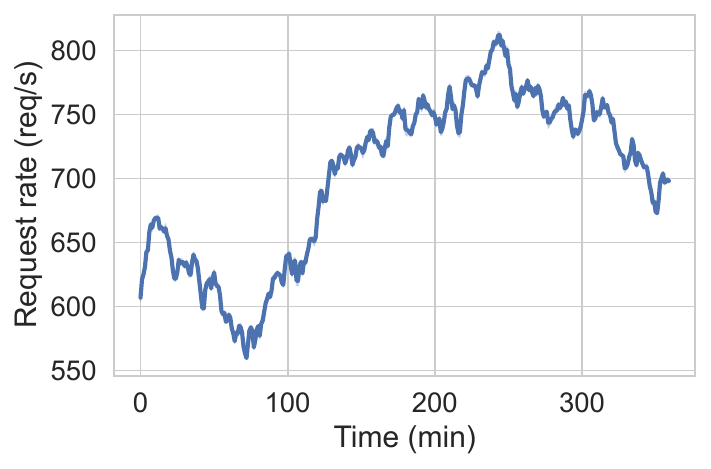}
        \caption{The Twitter streaming workload in a 6-hour window.}
    \end{subfigure}\hfill
    \begin{subfigure}{0.32\textwidth}
        \centering
        \includegraphics[width=\linewidth]{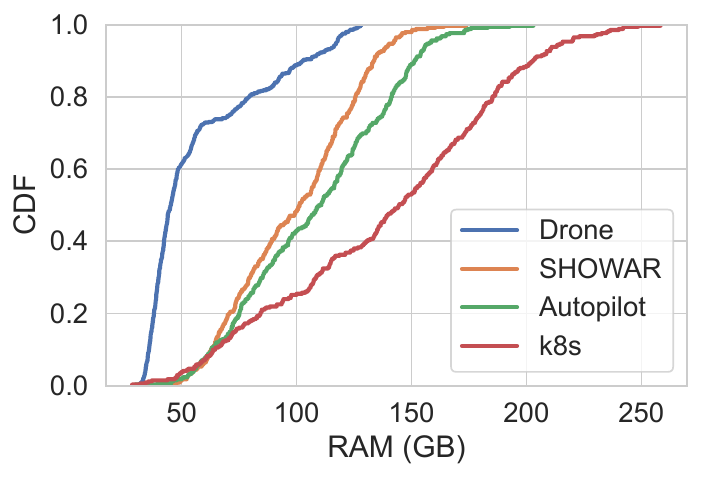}
        \caption{CDF of overall RAM allocation for SocialNet under the public cloud.}
    \end{subfigure}\hfill
    \begin{subfigure}{0.32\textwidth}
        \centering
        \includegraphics[width=\linewidth]{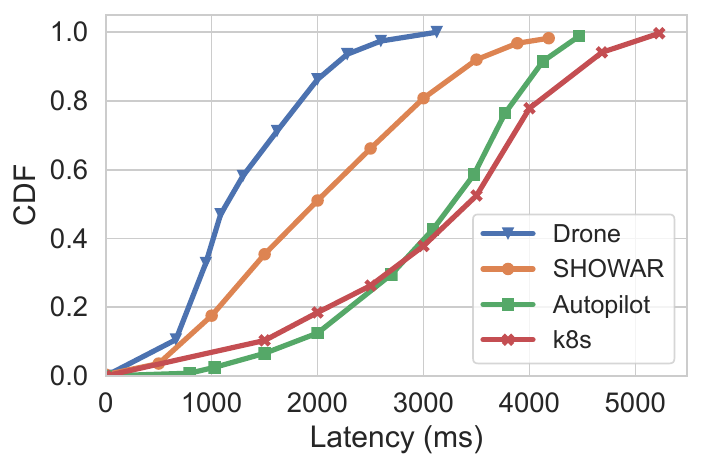}
        \caption{CDF of end-to-end latency for SocialNet under the public cloud.}
    \end{subfigure}
    \caption{Comparison between {\myName} and alternatives for microservices.\label{fig: eval_ms}}
\end{figure*}

Figure.~\ref{fig: eval_batch}(b) depicts the normalized resource cost saving compared to the Kubernetes native solution across all three representative batch workloads. While all three frameworks show cost-saving benefits thanks to their cost-aware problem formulation, {\myName} is the most cost-efficient one with over 20\% cost savings across workloads, since it can more accurately search for the (near-)optimal resource configuration based on the information from both performance feedback and environment contexts, without need to over-allocate resources to maintain reasonable performance. Moreover, {\myName} makes its own scheduling decision by incorporating the scheduling sub-vector into its action space. Thus, even if given the same total amount of resources, {\myName} also learns the best strategy to assign the execution pods to computing nodes, which Cherrypick and Accordia cannot achieve. This effect is most evident when benchmarking PageRank, a network-intensive workload, where {\myName} achieves an average of 53\% resource cost saving compared to the Kubernetes native solution, a number significantly higher than 20\% from Accordia and 17\% from Cherrypick.

Similar benefits are also manifested in the resource-limited private cloud setting. We focus on {\myName}'s impact on memory limit compliance since memory is a \textit{non-negotiable} resource type. Unlike CPU and network bandwidth where inadequate allocation would cause throttling (for CPU) or congestion (for network) but applications are still available, an application that requires more memory than allocated will incur an out-of-memory (OOM) error and the hosting pod will simply be killed and rescheduled if possible. OOM errors can significantly jeopardize application availability and degrade application performance. In our preliminary experiments in Sec.~\ref{obs}, a Spark job with insufficient memory allocation can experience a 20x longer elapsed time and even get halted in an intermediate stage and fail to make progress. We set the memory limit as 65\% of the total memory capacity available in the cluster and run all three representative batch workloads and record the memory utilization metric as shown in Figure.~\ref{fig: eval_batch}(c). We can observe that only {\myName} manages to abide by the memory constraint in the long run, showing an approximately 16\% lower memory profile than baselines, despite the first few exploration rounds where {\myName} actively explores around to identify the feasible safe action space. To see the benefit of resource limit compliance in action, we run memory-stressing tasks in parallel using \texttt{stress-ng} to simulate significant resource contention which occupies around 30\% of total memory. Table~\ref{tab:batch_private} summarizes the performance and number of Spark executor errors in different settings. We can observe that the Kubernetes native solution suffers from the least number of OOM errors by using memory utilization as one of its scaling rules. Therefore, it always respects the resource constraints and even suspends invoking executor pods when it detects memory is under stress, which in part contributes to its low performance. Memory constraint-oblivious solutions Cherrypick and Accordia on the other hand experience a large number of executor errors, especially for memory-intensive jobs such as LR and PageRank. In this case, {\myName} is able to fully utilize the algorithmic effectiveness of contextual bandits to optimize performance while complying with resource constraints, achieving up to 36\% performance improvement and 10x less OOM errors compared to Cherrypick and Accordia.


\subsection{{\myName} for Microservices}

We also evaluate the efficacy of {\myName} to orchestrate resources for microservice applications by performing end-to-end experiments. Driving the SocialNet microservice benchmark with a realistic workload trace as shown in Figure~\ref{fig: eval_ms}(a), we collected aggregated performance metrics over the entire application running span. Figure~\ref{fig: eval_ms}(b) shows the cumulative distribution of RAM allocation for {\myName} and the other three baselines. As hybrid autoscalers, both SHOWAR and Autopilot are able to reduce memory footprint compared to Kubernetes HPA by combining vertical and horizontal autoscaling to mitigate over-allocation. However, {\myName} outperforms the alternatives by more accurately modelling the performance-action relationship by incorporating a much broader array of factors, instead of only heavily relying on past resource usage information as SHOWAR and Autopilot do. Specifically, {\myName} is able to serve around 60\% of user requests within 50GB of overall RAM allocation, which is 55\% less than SHOWAR and 60\% less than Autopilot, manifesting an outstanding resource-saving capability.

Figure~\ref{fig: eval_ms}(c) depicts the end-to-end latency distribution across frameworks. Autopilot exhibits a similar performance to Kubernetes HPA since they share a similar reactive scaling strategy based on recent resource statistics. Specifically designed for microservices, SHOWAR performs better by identifying microservice correlations to create locality-oriented affinity rules, which is more likely to schedule closely related microservices to the same node and hence reduces latency. {\myName}, on the other hand, steps further by encoding the more efficient scheduling opportunities into its decision vector, so it does effectively both rightsizing (i.e., autoscaling by prioritizing vertical scaling) and scheduling. As an integrated resource orchestration solution, {\myName} lowers the P90 latency by 37\% compared to SHOWAR and by 45\% compared to Autopilot.

\begin{table}[t]
\begin{adjustbox}{width=\linewidth}
\begin{tabular}{|c|c|c|c|l|}
\hline
                      & k8s              & Autopilot        & SHOWAR           & {\myName} \\ \hline
\# of dropped packets & $4.8\times 10^4$ & $3.4\times 10^4$ & $1.4\times 10^4$ & 7809  \\ \hline
\end{tabular}
\end{adjustbox}
\caption{Number of dropped requests.}
\label{tab:drop}
\vspace{-0.7cm}
\end{table}

Having run the experiment in the private cloud setting, we also observe a similar effect as in the previous subsection. Table~\ref{tab:drop} records the total number of dropped user requests over the running span. Unlike in the batch processing case where the Kubernetes solution manages to maintain a low error rate by not invoking pods when memory is low, in this user-facing microservice case, it experiences the largest number of packet drops due to poor resource allocation decisions. Again, {\myName} incurs the least number of packet drops thanks to its resource limit-aware algorithm which progressively learns about the safe set for resource orchestration decisions.

\section{Discussion}\label{sec:discuss}

\noindent \textbf{Application-level configuration tuning.} An application's performance can greatly depend on its own configurable parameters as well. For example, Xin et al.~\cite{xin2022locat} identifies 38 performance-deciding parameters for SparkQL. While {\myName} is not readily applicable to application-level parameter tuning out of the box, the idea of the underlying contextual bandits as an algorithmic foundation can be naturally transferable, which has also been recently explored in the database research community, as mentioned in Sec.~\ref{sec:2.2}. In fact, {\myName} operates at a lower level of hardware resource orchestration and can be used in parallel with other efficient application-level configuration tuning techniques to jointly optimize an application's performance.

\noindent \textbf{Tradeoff between precision and cost.} In theory, the function modelling capability of bandit algorithms can always benefit from more information. It is also true in our resource orchestration context. Incorporating more dimensions for tunable parameters from the action space such as disk I/O and last level cache (LLC) bandwidth, or more contextual information such as the graph structure of the running microservices would help {\myName} more accurately characterize the complex coupling of performance, action and environment. However, it is well-known that bandit algorithms (especially its continuous variant Bayesian Optimization) tend to perform poorly in high dimensions~\cite{moriconi2020high}. Therefore, in practice, we need to selectively incorporate the ``more important'' dimensions with domain knowledge and employ several optimizations (see Sec.~\ref{sec:opt}) to make the algorithms practically efficient. Actually, as different applications and workloads have divergent resource request profiles, it would be interesting to investigate how to dynamically pick the most critical dimensions based on application and workload properties. We leave that as a future work of {\myName}.

\noindent \textbf{Overhead of {\myName}.} {\myName} is designed to embrace the latest cloud paradigms and technologies, working seamlessly with the Kubernetes ecosystem. It utilizes the Prometheus-based Kubernetes monitoring stack for metrics collection and modifies resource configurations by directly communicating with the Kubernetes API server and updating the \texttt{cgroup} configuration values for the pods of concern if possible. Thanks to the optimizations we employ, the computation time for each iteration in the online mode is on the order of seconds, well within the metrics updating interval. There is also no additional cost of potential container migration during scheduling since {\myName} follows the standard Kubernetes-native rolling-update procedure. Therefore, minimal overhead is incurred for using {\myName}. 

\noindent \textbf{Limitations.} One major limitation of {\myName} is its insufficient capability to deal with ``flash crowds'', workloads that burst to a significantly higher level in a very short period of time (e.g., seconds). This situation inherently breaks the Gaussian Process prior assumption of the function and intrinsic limitations of iterative algorithms restrict {\myName} from reacting fast to such sudden changes. Fortunately, such cases are rare in reality and cloud providers often prepare backup resources for over-allocation in addition to their routine resource allocation frameworks. Moreover, {\myName} is yet to achieve its full potential to work with microservices since it is oblivious to the microservice dependency graph structure, which has been shown to be instrumental in microservice-oriented resource management~\cite{qiu2020firm,luo2021characterizing,baarzi2021showar,luo2022power,park2021graf}. Efficiently integrating dependency information into {\myName} without incurring significant overhead would be another promising direction to explore.

\section{Conclusions}\label{conclusion}

In this paper, we present {\myName}, a resource orchestration framework specifically designed for the containerized cloud. Based on recent advances in contextual bandit algorithms, {\myName} encapsulates various cloud uncertainties as contextual parameters to aid the search process for optimal resource orchestration decisions. The uncertainty-aware approach enables {\myName} to progressively balance the performance and resource cost tradeoff in a shared public cloud, and optimize performance while adhering to resource constraints in a resource-limited private cloud. Our empirical analysis shows that {\myName} achieves up to 45\% performance improvement and 20\% resource cost savings compared to state-of-the-art alternatives.

\bibliographystyle{ACM-Reference-Format}
\bibliography{sample-base}

\end{document}